\documentclass[preprint,showpacs,preprintnumbers,amsmath,amssymb,prd,citeautoscript,epsf]{revtex4}
\usepackage{graphicx}
\usepackage{epsfig}

\begin{document}

\title{Conditions for detecting CP violation via neutrinoless double beta
 decay}
\author{A. \surname{Joniec}}
\author{M. \surname{Zralek}}
\affiliation{Institute of Physics, University of Silesia
\\Uniwersytecka 4, 40-007 Katowice, Poland}
\begin{abstract}
Neutrinoless double beta decay data together with information on
the absolute neutrino masses obtained from the future KATRIN
experiment and/or astrophysical measurements give a chance to
find CP violation in the lepton sector with
Majorana neutrinos. We derive and discuss necessary conditions
which make discovery of such CP violation possible for the future
neutrino oscillation and mass measurements data.
\end{abstract}
\pacs{14.60.Pq, 26.65.+t, 95.85.Ry}
 \maketitle
\section{Introduction}
Information on CP violation in the lepton sector is very
important for building the future theories which go beyond the Standard
Model \cite{Aalseth}. As CP violation is probably
predominantly connected with lepton masses and observed neutrinos
are very light, an experimental measurement of the
effect is a serious challenge. For three Dirac neutrinos there is one
CP violating phase $(\delta)$ and
additional two phases $(\alpha_1, \alpha_2)$ exist for Majorana
neutrinos. The charged current state $(\nu_{\alpha})$ is related to
mass states $(\nu_{i})$ by an unitary transformation
\begin{equation}
|\nu_{\alpha}\rangle=\sum_i U^*_{\alpha i}|\nu_i\rangle
\end{equation}
where
\begin{equation}\label{param}
\begin {array}{c}
 U_{\alpha i}=\left(
\begin{array}{ccc}
c_{12}c_{13} & s_{12}c_{13} & s_{13}e^{-i\delta} \\
-s_{12}c_{23}-c_{12}s_{23}s_{13}e^{i\delta} &
c_{12}c_{23}-s_{12}s_{23}s_{13}e^{i\delta} & s_{23}c_{13} \\
s_{12}s_{23}-c_{12}c_{23}s_{13}e^{i\delta} &
-c_{12}s_{23}-s_{12}c_{23}s_{13}e^{i\delta} & c_{23}c_{13}
\end{array}
\right) \times \\ \\ \left(
\begin{array}{ccc}
e^{i\alpha_1} & 0 & 0 \\
0 & e^{i\alpha_2} & 0 \\
0 & 0 & 1
\end{array}
\right).
\end{array}
\end{equation}
$c_{ij}$ and $s_{ij}$ are cosines and sines of the $\theta_{ij}\:
(ij=12, 13, 23)$ angles. The second matrix in (\ref{param}) appears
only for Majorana neutrinos.
\par It is commonly believed that CP violation owing to Dirac phase
$\delta$ will be discovered in the future superbeam or neutrino
factory experiments \cite{Geer,Geer2} where oscillations of
neutrinos and antineutrinos will be observed. From the
parametrization of the mixing matrix (Eq.\ref{param}) we can see that
$\sin\theta_{13}$ and $e^{\pm i\delta}$ always appear in a
combination. So, any CP breaking effect for Dirac neutrinos will
be proportional to $\sin\theta_{13}\sin\delta$ and disappear for
$\sin\theta_{13}\rightarrow 0$. From the present fits it follows that
this mixing angle is small ($\sin^2\theta_{13}<0.05$ for $99.7\%$
C.L. \cite{Bandy,CHOOZ,PaloVerde}) and the assumption that
$\theta_{13}=0$ agrees with the data equally well. Such a tendency,
if outlives in the future when more precise data will be available,
effects in a very small or
vanishing CP breaking signal. It was shown that for
$\delta=\pm\frac{\pi}{2}$ effects of CP violation will be seen in
the future experiments if $\sin^2\theta_{13}$ is not smaller than
$10^{-4}$ \cite{Lindner}.
\par If neutrinos are Majorana particles, in addition to the phase
$\delta$, two other phases can also be responsible for CP symmetry
breaking. Many different processes are, in principle, sensitive to
these Majorana phases and can generate both CP-even and CP-odd
effects \cite{Matsuda,Gouvea}. Admittedly, most of them are much
beyond an observable limit. The only experiment
which could provide evidence for Majorana phases is the
search for neutrinoless double beta decay $(\beta\!\beta)_{0\nu}$.
Such a possibility has been discussed  many times
\cite{Barger2,Barger3,Bilenky,Bilenky2,Bilenky3,Czakon,Czakon2,Falcone,Klapdor,Osland2,Minakata,Vissani,Rodejohann,Abada}
but, to our knownledge, detailed conditions concerning the future
experimental results and
their necessary precision to discover CP violation has not been
discussed. The exception is where Ref. \cite{Barger4} authors consider the
future anticipated precision of all relevant neutrino
experiments, and formulate a very pessimistic 'no-go' conclusion.
They state that even under a very optimistic assumption about
sensitivity of the future experiments it will be impossible to detect
neutrino CP violation in the $(\beta\!\beta)_{0\nu}$ decay. We agree
with such a statement, but we would like to go a step further.
 We propose a
set of conditions for neutrino masses and mixing angles (best fit
values = b.f.v) altogether with conditions on experimental and theoretical
precision for
their determination, such that discovery of CP violation arising
from Majorana phases in the $(\beta\!\beta)_{0\nu}$ decay will be
possible. We formulate sufficient conditions when CP violation
could be observed. We should mention that our conditions are completely
general. Contrary to Ref. \cite{Barger4} we do not assume from the
beginning that the $\theta_{13}$ angle vanishes. Similar
consideration has been done in \cite{Pascoli,Pascoli3,Petcov}. Here we
concentrate on the degenerate neutrino mass spectrum where CP
violation has a clear meaning. We investigate in more details the
problem of theoretical determination of the nuclear matrix
elements, mechanism responsible for $(\beta\beta)_{0\nu}$ and the
future experimental error of nuclei decay lifetime.
\par We found that under a very optimistic assumption on the sensitivity of
the future experiments considered in Ref. \cite{Barger4}, independently
of measured b.f.v., it is really impossible to detect CP
violation. However, such a possibility is 'just around the corner'.
A little better precision will give a chance to make a decisive
statement about CP Majorana breaking. Even if required precision for today
is estimated to be very optimistic value, we  hope that the problem of
lepton CP violation is so important that it is worth to have it in
mind.
\par Other important result of our investigation concerns the $\theta_{13}$ and
$\theta_{12}$ mixing angles. In contrary to neutrino oscillation
experiments, smaller $\theta_{13}$ angle gives better prospect of
CP symmetry breaking measurement. Similar situation take place for solar mixing
angle $\theta_{12}$. The maximal $\theta_{12}$,
$\sin2\theta_{12}\rightarrow1$, is the best scenario for CP
symmetry breaking discovery, contrary to the prospects of finding
the neutrino mass bound from $(\beta\beta)_{0\nu}$ decay
\cite{Joniec}, where $\theta_{12}\rightarrow\frac{\pi}{4}$ brings
such a possibility to ruin.
\par In the next chapter we discuss how CP symmetry braking could be
determined from neutrinoless double beta decay. Then, in Chapter
III, we describe the present situation  and we predict how
precisely all parameters (oscillation mixing angles, effective
mass $\langle m_\nu\rangle$ measured in $(\beta\beta)_{0\nu}$ and
$m_\beta$ measured in e.g. tritium beta decay) should be determined
in order to discover CP symmetry breaking. Two kinds of
presentations are given. The first one, very visual, where
correlations between errors are not included. And the second,
where more sophisticated analysis which answer at which value of
confidence level the probes of CP violation could be carried on. Finally,
Chapter IV contains our conclusions.

\section{CP symmetry breaking and the $(\beta\beta)_{0\nu}$ decay.}

\par
The neutrinoless double beta decay  $(\beta\beta)_{0\nu}$ of nuclei measures
the effective neutrino mass $\langle m_{\nu}\rangle$ \cite{Doi}:
\begin{equation}\label{M_mass}
\langle m_{\nu}\rangle=|\sum_{i=1}^3 U_{ei}^2 m_i|=|c_{12}^2
c_{13}^2 m_1 + s_{12}^2 c_{13}^2 m_2 e^{2i\phi_2} + s_{13}^2 m_3
e^{2i\phi_3}|,
\end{equation}
where $\phi_2=\alpha_2-\alpha_1$ and $\phi_3=-\delta-\alpha_1$.
\par As we will see, the possible precision of the future experiments will give
a chance to look for CP violation only for higher neutrino masses
$(m_1\gtrsim 0.1\: eV)$. For this case the mass spectrum starts to be
degenerated and we will consider only such a spectrum. Then
the effective neutrino mass $m_{\beta}$ measured in tritium beta
decay, independently of its definition \cite{Alberico}, is just
equal to neutrino masses
\begin{equation}
 m_{\beta}= \left[\sum^3_{i=1} |U_{ei}|^2 m^2_i\right]^{1/2}=\sum^3_{i=1} |U_{ei}|^2
 m_i = m_1\approx m_2 \approx m_3.
\end{equation}
For Majorana neutrinos CP symmetry holds if $\alpha_i$ and $
\delta$ take one of the values $0, \pm \frac{\pi}{2}, \pm \pi$.
Then from Eq. (\ref{M_mass}) four conserving CP values of $\langle
m_{\nu}\rangle$ are obtained:
\begin{eqnarray}
 \langle m_{\nu}\rangle_{(1)} & = & m_{\beta}, \nonumber \\
 \langle m_{\nu}\rangle_{(2)} & = & m_{\beta}\cos 2\theta_{13}, \nonumber \\
 \langle m_{\nu}\rangle_{(3)} & = & m_{\beta}\left(\cos^2\theta_{13}|\cos2\theta_{12}|+
 \sin^2\theta_{13}\right), \\
 \langle m_{\nu}\rangle_{(4)} & = & m_{\beta}\left(\cos^2\theta_{13}|\cos2\theta_{12}|-
\sin^2\theta_{13}\right). \nonumber
\end{eqnarray}
In all cases, the relation between $ \langle m_{\nu}\rangle$ and
$m_{\beta}$ is linear with different slopes $(i=1,2,3,4)$
\begin{equation}\label{linear}
 \langle m_{\nu}\rangle_{(i)}=c_i m_{\beta}.
\end{equation}
First we would like to present very visual method of finding a region of
parameters where CP violation can be probe. Lately we will present a result
with the correct statistical analysis.
Let us assume that $\theta_{ij}$ mixing angles are known with definite precision,
\begin{equation}
\sin^2\theta_{ij}\in ((\sin^2\theta_{ij})_{min}, (\sin^2\theta_{ij})_{max})
\end{equation}
with central value
\begin{equation}
(\sin^2\theta_{ij})_{best fit}.
\end{equation}
For each $c_i\; (i=2,3,4)$ we can calculate the maximal and minimal values
\begin{eqnarray}
c^{max}_2=(\cos 2\theta_{13})_{max},  \nonumber \\
c^{min}_2=(\cos 2\theta_{13})_{min},  \nonumber \\
c^{max}_3=(\cos^2\theta_{13})_{max}(\cos 2\theta_{12})_{max}+
(\sin^2\theta_{13})_{max}, \nonumber  \\
c^{min}_3=(\cos^2\theta_{13})_{min}(\cos 2\theta_{12})_{min}+
(\sin^2\theta_{13})_{min},   \\
c^{max}_4=(\cos^2\theta_{13})_{max}(\cos 2\theta_{12})_{max}-
(\sin^2\theta_{13})_{min},  \nonumber \\
 c^{min}_4=(\cos^2\theta_{13})_{min}(\cos 2\theta_{12})_{min}-
(\sin^2\theta_{13})_{max}. \nonumber
\end{eqnarray}
We can see that localization of the $\langle m_{\nu}\rangle_{(i)}$
lines is fully determined by the oscillation parameters, namely
$\theta_{13}$ and $\theta_{12}$ angles.
\begin{figure}
\includegraphics[width=\textwidth]{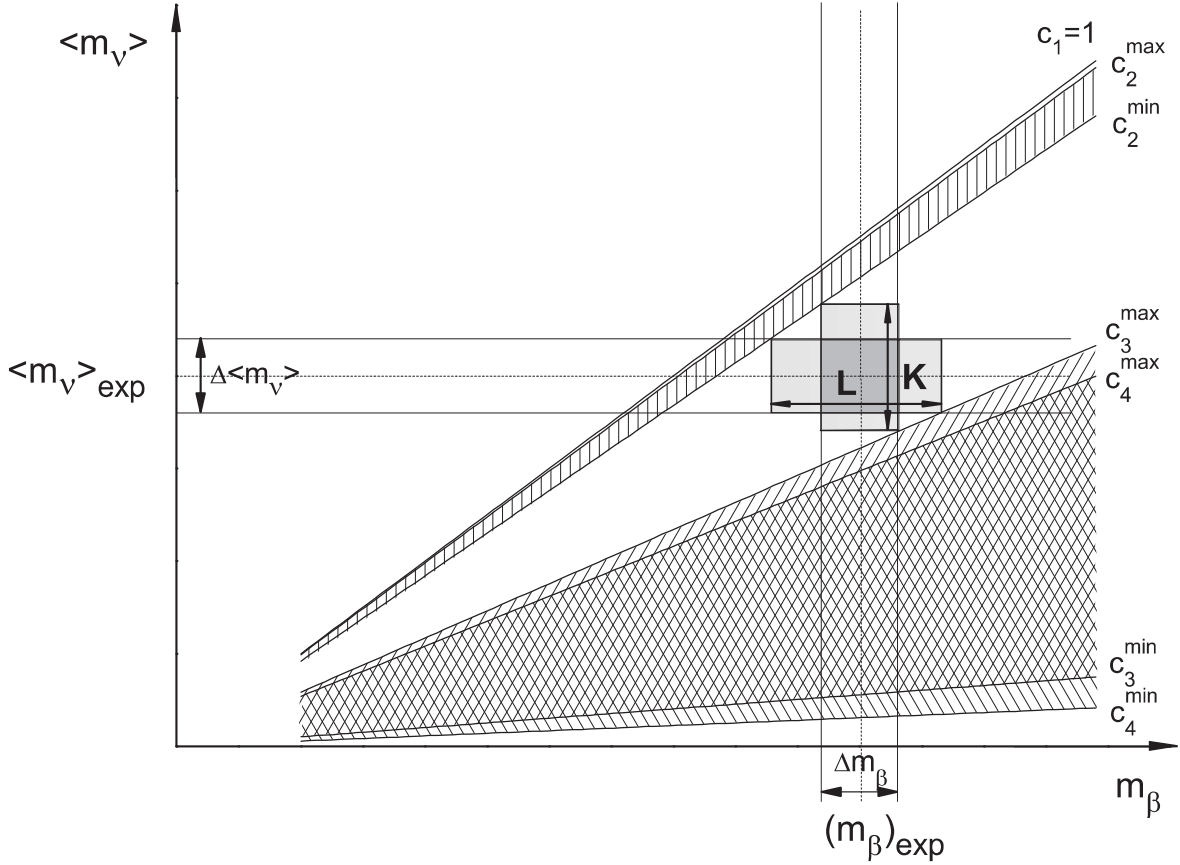}
\caption{A localization of the $R=\left(\Delta m_\beta ,\Delta
\langle m_\nu\rangle \right)$ rectangle between $c_2^{min}$ and
$c_3^{max}$ lines which indicates that CP symmetry is broken.}
\label{fig}
\end{figure}
\par Let us now assume that in the future experiments $m_{\beta}$
and $\langle m_{\nu}\rangle$ masses are determined with precision
$\Delta m_{\beta}$ and  $\Delta \langle m_{\nu}\rangle$:
\begin{eqnarray}
\langle m_\nu\rangle_{exp}&\pm&\Delta\langle m_\nu\rangle, \\
(m_\beta)_{exp}&\pm&\Delta m_\beta.
\end{eqnarray}
Then localization of the rectangle $R=(\Delta
m_\beta,\Delta\langle m_\nu\rangle)$ between the lines $c_1=1$ and
$c_4^{min}$ (see Fig.\ref{fig}) decides about CP symmetry
breaking. If $R$ crosses the error region between the
$(c_i^{min}, c_i^{max})$ lines $i=2,3,4$, we do not know anything
about CP symmetry. But, in opposite, if $R$ is located outside the
$c_i$ error region then there is indication that CP symmetry is
broken as at least one of the angles $\delta, \alpha_1,\alpha_2$
is not equal to its CP conserving value.
\begin{figure}
\includegraphics[width=\textwidth]{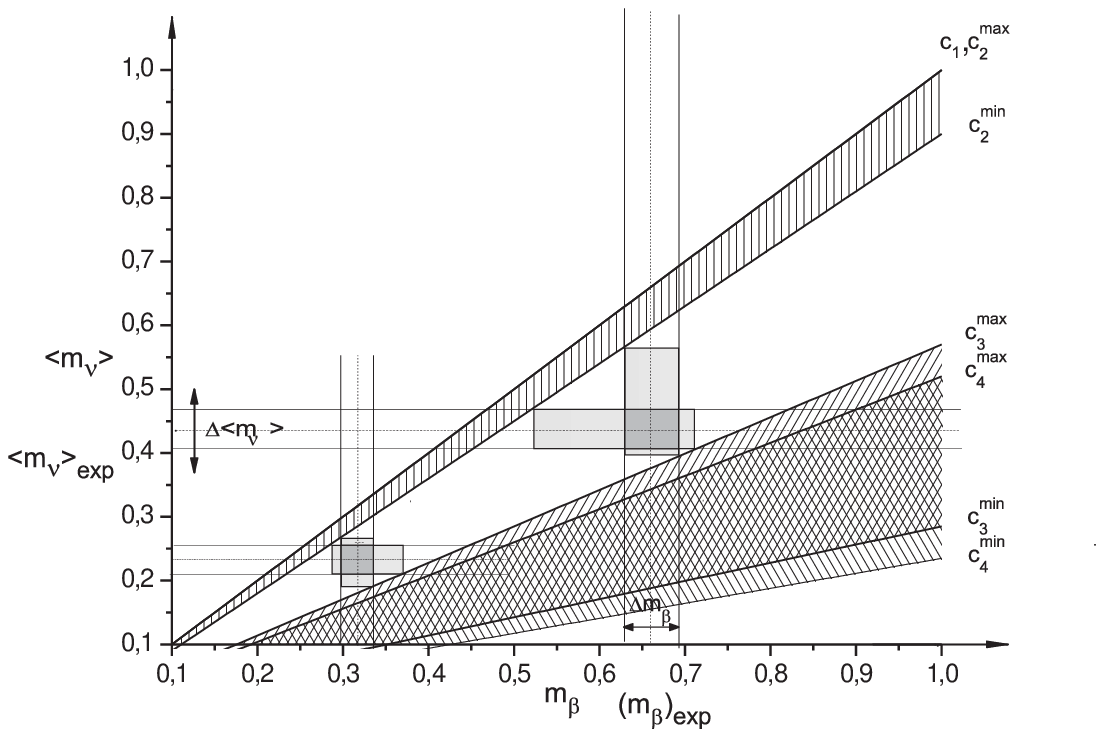}
\caption{A localization of the ($c_i^{min}\; c_i^{max})$ regions for
the present $\theta_{13}$ and $\theta_{12}$ angles precision. To
see CP violation a precision of $m_\beta$ and $\langle
m_\nu\rangle$ measurements should be very good. For smaller
$m_\beta$ (and $\langle m_\nu\rangle$) a region where CP
violation can be search for is smaller, so a precision of their
measurements should be even better.} \label{present}
\end{figure}
\begin{figure}
\includegraphics[width=\textwidth]{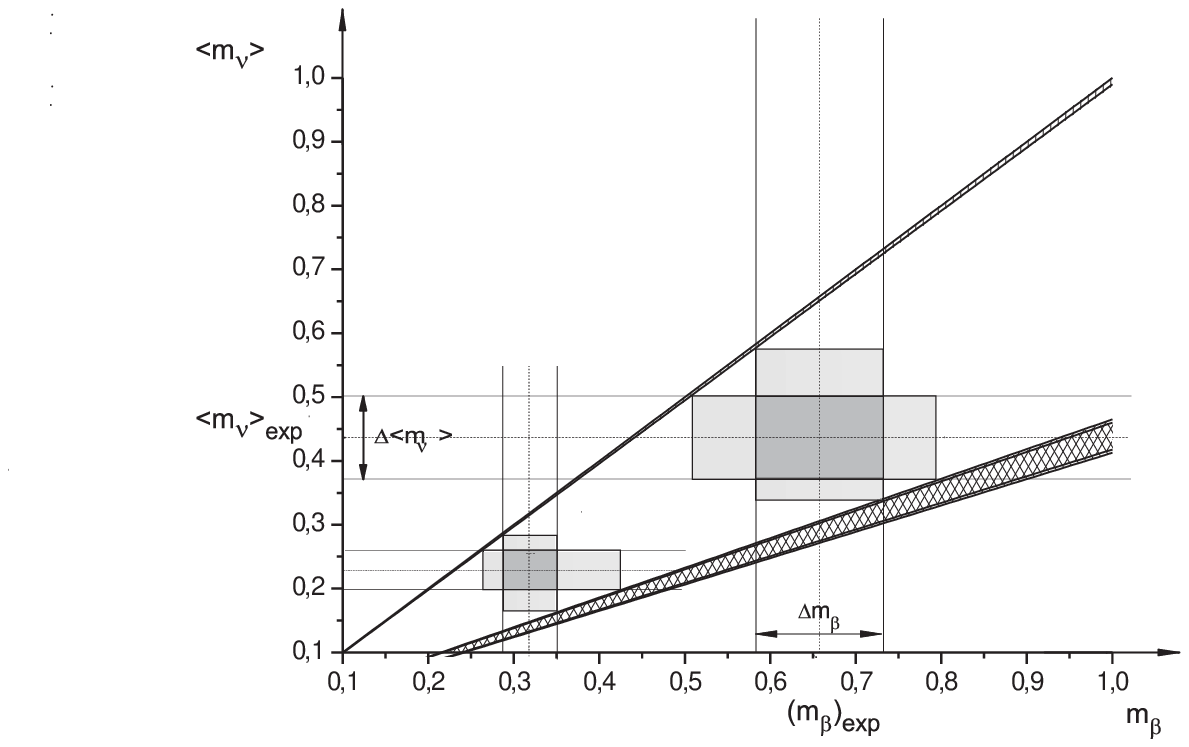}
\caption{The CP conserving regions (hatched areas) which follow from
the future neutrino oscillation experiments. We assume that
central values of $\theta_{13}$ and $\theta_{12}$ are in agreement with present data but their error estimation is suppose to be much better
($\sin^2\theta_{12}=0.28\pm 0.01$ and $\sin^2\theta_{13}=0.005\pm
0.0001$). The region between $c_2^{min}$ and $c_3^{max}$ lines is
larger, giving more space for the rectangle ($\Delta m_\beta,\;\Delta\langle
m_\nu\rangle$) (see text for more details).
}\label{future}
\end{figure}
\par Possible localization of the present and prospective $\langle
m_\nu\rangle_i=c_i m_\beta$ lines is presented in Fig.
\ref{present} and \ref{future}, respectively. We can see that
localization of $R$ between $c_3^{max}$ and $c_2^{min}$ lines is
only interesting for CP violation search. If the rectangle $R$
with $\Delta m_\beta$ and $\Delta\langle m_\nu\rangle$ sides is
fully located between two lines with the $c_3^{max}$ and
$c_2^{min}$ slopes then CP symmetry is broken (see Fig.\ref{fig}).
So first conditions for detecting CP violation are:
\begin{equation}\label{KL}
 \Delta m_{\beta}<L,\; \; \; \; \; \; \Delta\langle
 m_{\nu}\rangle<K.
\end{equation}
L and K can be find in a easy way
\begin{equation}\label{K}
 K= (m_{\beta})A-(\Delta m_{\beta}) B,
\end{equation}
and
\begin{equation}\label{L}
 L=\langle m_{\nu}\rangle C-\Delta \langle m_{\nu}\rangle D,
\end{equation}
where
\begin{eqnarray}
 A=c_2^{min}-c_3^{max}, \nonumber \\
 B=\frac{c_2^{min}+c_3^{max}}{2}, \nonumber \\
 C=\frac{A}{c_2^{min}c_3^{max}}, \label{ABCD}\\
 D=\frac{B}{c_2^{min}c_3^{max}}. \nonumber
\end{eqnarray}
for any $m_{\beta}$ and $\langle m_{\nu}\rangle$ values inside the two lines
$c_2^{min}$ and $c_3^{max}$.
\par If conditions (Eq. \ref{KL}) are satisfied for some central values
$(m_{\beta})_{exp}$ and $\langle m_{\nu}\rangle_{exp}$ determined
from experiments (and theory) then there are further two possibilities. The
rectangle $R$ located in the point $\left((m_{\beta})_{exp},
\:\langle m_{\nu}\rangle_{exp}\right)$ can
\begin{enumerate}
 \item be fully inside two bounding lines $c_2^{min}$ and $c_3^{max}$, or
 \item be located partly on the first or the second line.
\end{enumerate}
In the first case we can conclude that CP symmetry is broken, in
the second the problem is unresolved. The first condition is
satisfied if:
\begin{equation}\label{inside}
c_3^{max}\left( (m_{\beta})_{exp}+\frac{\Delta
m_{\beta}}{2}\right)<\left(\langle
m_{\nu}\rangle_{exp}-\frac{\Delta\langle
m_{\nu}\rangle}{2}\right),
\end{equation}
and
\begin{equation}\label{inside2}
\left(\langle m_{\nu}\rangle_{exp}+\frac{\Delta\langle
m_{\nu}\rangle}{2}\right)<\left( (m_{\beta})_{exp}-\frac{\Delta
m_{\beta}}{2}\right)c_2^{min}.
\end{equation}
The inequalities given by Eqs.(\ref{KL}), (\ref{inside}) and
(\ref{inside2}) form the set of necessary conditions for CP
symmetry breaking. Of course we are not able to
prove in this way that CP symmetry holds.
\par Let us parameterize
\begin{equation}
\Delta\langle m_{\nu}\rangle=2x\langle m_{\nu}\rangle, \; \; \; \;
\Delta m_{\beta} = 2y\, m_{\beta},
\end{equation}
where $2x$ is the relative error which measures the theoretical
nuclear matrix elements uncertainty and experimental decay lifetime of the $\langle m_\nu\rangle $ matrix element.
Similarly, $2y$ measures the
relative error of the effective mass e.g. from tritium beta decay.
As both $K$ and $L$ (in Eq. (\ref{K}) and (\ref{L})) must be
larger than zero, we have two consistency conditions. Both $x$ and
$y$ must satisfy the same inequality
\begin{equation}\label{xy}
x,\, y \leq \frac{1-\cos2\theta_{12
\mathrm{\;\mathrm{min}}}-3\sin^2\theta_{13\;
\mathrm{\;\mathrm{max}}}+ \sin^2\theta_{13\;
\mathrm{min}}\cos2\theta_{12\; \mathrm{min}}} {1+\cos2\theta_{12\;
\mathrm{min}}-\sin^2\theta_{13\; \mathrm{max}}-\sin^2\theta_{13\;
\mathrm{min}}\cos2\theta_{12\; \mathrm{min}}}.
\end{equation}
These inequalities impose sharp conditions concerning a precision
of the $m_{\beta}$ and $\langle m_{\nu} \rangle$ determination. As
the r.h.s. of Eq. (\ref{xy}) is a decreasing function of
$\sin^2\theta_{13}$ and $\cos2\theta_{12}$, the best circumstances
arise for $\sin^2\theta_{13}\rightarrow 0$ and $\sin^2\theta_{12}
\rightarrow \frac{1}{2}$. In this case lines $c_2^{min}
\rightarrow 1$ and $c_3^{max}\rightarrow 0$ give the largest
region for localization of $\langle m_{\nu}\rangle_{exp}$ and
$(m_{\beta})_{exp}$ where symmetry is broken. As we know, the
condition $\theta_{13}\rightarrow 0$ ruins the Dirac $\delta$
phase determination in oscillation experiments. We can see that both
methods, $(\beta\beta)_{0 \nu}$ decay and long baseline experiment
which could detect $\delta$ are complementary for detecting CP
violation \cite{Pascoli2}. Also the other condition, the large solar
mixing angle ($\theta_{12}\rightarrow\frac{\pi}{4}$) is not
favourable for Majorana mass determination from the
$(\beta\beta)_{0\nu}$ decay [7-18].
\par The case $\sin^2\theta_{13}=0$ has been considered in Ref.\cite{Barger4}.
Then Eq.(\ref{xy}) gives
\begin{equation}\label{mixang}
x<\tan^2\theta_{12}
\end{equation}
which is exactly the condition given by Eq. (14) in
Ref.\cite{Barger4}.
\par From Eqs. (\ref{K}) and (\ref{L}) for given
relative errors $x$ and $\Delta m_\beta $ we can
also find the lower limit for the $m_\beta$ and $\langle
m_\nu\rangle $ effective masses for which measurements  are still possible
\begin{equation}\label{lowlimn}
\langle m_\nu\rangle > \frac{\Delta m_\beta}{C-2xD},
\end{equation}
and
\begin{equation}\label{lowlimb}
m_\beta > \frac{\Delta m_\beta}{A}\left(B+\frac{2x}{C-2xD}\right).
\end{equation}
Now, using present and expected in the future precision of the neutrino
oscillation data we can estimate how well  $m_\beta$ and
$\langle m_\nu\rangle$ should be determined to discover CP
symmetry breaking.
\section{Numerical results}
Using presently determined $\theta_{12}$ and $\theta_{13}$
mixing angles \cite{Bahcall,Gonzales,Goswami,Maltoni2,Fogli2} (with
$3\sigma$ precision)
\begin{eqnarray}\label{data}
0.22\leqslant & \sin^2\theta_{12} & \leqslant0.37, \nonumber \\
0\leqslant & \sin^22\theta_{13} & \leqslant 0.048,
\end{eqnarray}
from Eq.(\ref{xy}) we obtain:
\begin{equation}\label{x-nes}
x<0.2.
\end{equation}
It will be a serious challenge to get such a precision. Let us check
it for the isotope of Germanium $^{76}Ge$ where evidence for the
$(\beta\beta)_{0\nu}$ decay is claimed to have been obtained
\cite{Klapdor2}. If we assume that only one standard mechanism,
the exchange of Majorana neutrinos with masses $m_i$, is
responsible for the $(\beta\beta)_{0\nu}$ decay, the effective mass
$\langle m_\nu\rangle$ is calculated from decay rate $T(^{76}Ge)$
\cite{Elliott}:
\begin{equation}\label{lifetime}
T^{-1}(^{76}Ge)=G|M|^2\langle m_\nu\rangle^2,
\end{equation}
where $G$ is accurately calculable phase space integral and
$M$ is calculated Nuclear Matrix Element (NME). Unfortunately, this
calculation is a complicated job, and different methods of
calculation give different results. For the isotope $^{76}Ge$ the
results differ by one order of magnitude. If we parameterize
\begin{equation}
T(^{76}Ge)=b\times 10^{24}\;\mathrm{y},
\end{equation}
then for $\langle m_\nu\rangle=1$ eV eleven different results have
been obtained
\cite{Joniec}:
\begin{figure}
\includegraphics[width=\textwidth]{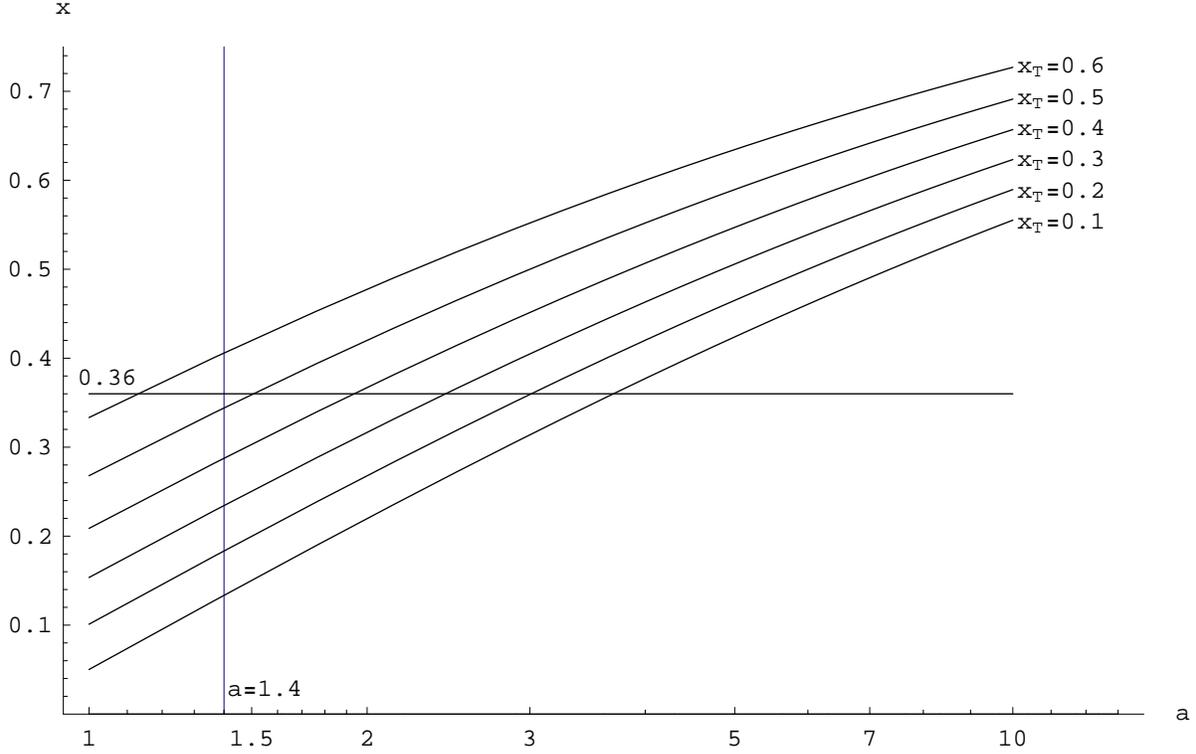}
\caption{The lines for the full uncertainty of $\langle
m_\nu\rangle$, $x=\frac{\Delta\langle m_\nu\rangle}{2\langle
m_\nu\rangle}$ as a function of both, theoretical uncertainty in
 NME calculations $a=\frac{b_{max}}{b_{min}}$ (see text) and
experimental relative error $(x_T)$ for the decay lifetime of
$^{76}Ge$.} \label{fig4}
\end{figure}
\begin{center}
\begin{tabbing}
  cccccccccccccccccccccc \= cccccccccccc \= cccccccccccc \= cccccccccccc \= cccccccccccc \= cccccccccccc  \kill
  $b=1.7 (=b_{min})$ \cite{Haxton1}, \> 2.16 \cite{Tomoda}, \> 2.3 \cite{Engel}, \>
  2.33 \cite{Staudt}, \> 3.15 \cite{Feassler}, \> 3.2 \cite{Feassler}, \\ 3.6
  \cite{Pantis},
  \> 4.06 \cite{Suhonen}, \> 8.95 \cite{Simkovic}, \> 14.0 \cite{Engel1}, \> 17.7 $(=b_{max})$ \cite{Caurier}. \>
\end{tabbing}
\end{center}
However, we would like to stress, that method used in Ref.
\cite{Haxton1,Tomoda, Engel, Staudt, Feassler, Pantis, Suhonen, Simkovic, Engel1, Caurier}
are completely independent, different nuclear models are used, and generally
models are not calibrated against nuclear properties.
If we assume that relative experimental error for $T(^{76}Ge)$
measurements is defined by $2x_{T}$
\begin{equation}
 \frac{\Delta T(^{76}Ge)}{\langle T(^{76}Ge)\rangle}=2x_T,
\end{equation}
then the full relative uncertainty of $\langle m_\nu\rangle$
($2x=\frac{\Delta \langle m_\nu\rangle}{\langle m_\nu\rangle}$) is
given by ($x_T<1$)
\begin{eqnarray}\label{x}
x=x(a,x_T)=\frac{\sqrt{(1+x_T)a}-\sqrt{1-x_T}}{\sqrt{(1+x_T)a}+\sqrt{1-x_T}},
\end{eqnarray}
where $a=\frac{b_{max}}{b_{min}}$. In Fig.4 we can see the relation
between the NME precision $(a)$ and the expected uncertainties for
effective neutrino mass $\langle m_\nu\rangle$ $\left(
x=\frac{\Delta \langle m_\nu\rangle} {2\langle m_\nu\rangle}
\right)$ for various future experimental errors of the decay lifetime
of $^{76}Ge$.
\par We can see, that taking seriously present discrepancy in the NME determination
 $(a\thickapprox 10)$ we obtain $x\thickapprox 0.52$, much larger than necessary
 (see Eq.(\ref{x-nes})).
The new calculation of NME \cite{Rodin}, where the observed
$(\beta\beta)_{2\nu}$ decay has been used to fix relevant
parameters, has shown the great stability of the final results.
For the $^{76}Ge$ two methods of calculation,
RQRPA and QRPA (see \cite{Rodin} for more details) have given
almost the same results, and then:
\begin{equation}
a\thickapprox 1.4.
\end{equation}
With  such a precision of the NME determination we obtain ($x_T\thickapprox 0.3$)
\begin{equation}
x\thickapprox 0.24,
\end{equation}
still above the present necessary precision (see Eq.(\ref{x-nes})), but within
reach of the future oscillation experiments.
\par We should also mention the other
uncertainty in the $\langle m_\nu \rangle$ determination - the
possible different physical mechanism for the
$(\beta\beta)_{0\nu}$ decay. If the lepton number is violated at TeV
scale we can expect the other processes which give equally strong,
as light Majorana neutrinos exchange, contributions to
$(\beta\beta)_{0\nu}$. Then the relation between decay lifetime
and $\langle m_\nu\rangle$ is not given by
Eq.(\ref{lifetime}). To answer a question at which scale
lepton number is violated,  information from higher energy
colliders (e.g. LHC) and other lepton processes is necessary. In
Ref. \cite{Cirigliano} it was shown that a study of
two lepton flavor violating processes $\mu\rightarrow e$
conversion and $\mu\rightarrow e +\gamma$ decay will give
important insight to the mechanism of the $(\beta\beta)_{0\nu}$ decay.
\par From Eqs.(\ref{lowlimn}) and (\ref{lowlimb}) we can find
conditions for $m_\beta$ and $\langle m_\nu\rangle$ effective
masses for which CP symmetry breaking could be seen (see
Fig(\ref{present})). For example, if $x\thickapprox 0.15$ with the
present $3\sigma$ precision of mixing angles Eq.(\ref{data}) and for
$\Delta m_\beta=0.03,0.02,0.015$ eV, the CP symmetry breaking is
testable for $\langle m_\nu\rangle>0.24, 0.16,0.12$ eV and
$m_\beta>0.32, 0.21,0.16$ eV, respectively. There is some chance
that in the future experiments such $\Delta m_\beta$ precision can
be reached, but relative error for $\langle m_\nu\rangle$,
$x\thickapprox0.15$ is much beyond the present possibilities.
\par From Eq. (\ref{lowlimn}) for a given central value of
$\langle m_\nu\rangle$, we can find relation between the $x$ and
$\Delta m_\beta$ required for probing the CP symmetry breaking.
Let us assume that a value of $\langle m_\nu\rangle$ is really
in an interval given by the Heidelberg group \cite{Klapdor2}
\begin{equation}\label{HM}
\langle m_\nu\rangle_{exp }\thickapprox(0.1-0.9)\; \mathrm{eV}.
\end{equation}
If $\langle m_\nu\rangle_{exp}\thickapprox0.1$ $(0.9)$ eV, $\Delta
m_\beta$ should be smaller than $0.002, 0.013, 0.026$ ($0.014,
0.11, 0.24$) eV for $x\thickapprox 0.19, 0.15$ and $0.1$,
respectively, with the central value $m_\beta\thickapprox0.13$
($1.2$) eV.
\begin{figure}
\includegraphics[width=\textwidth]{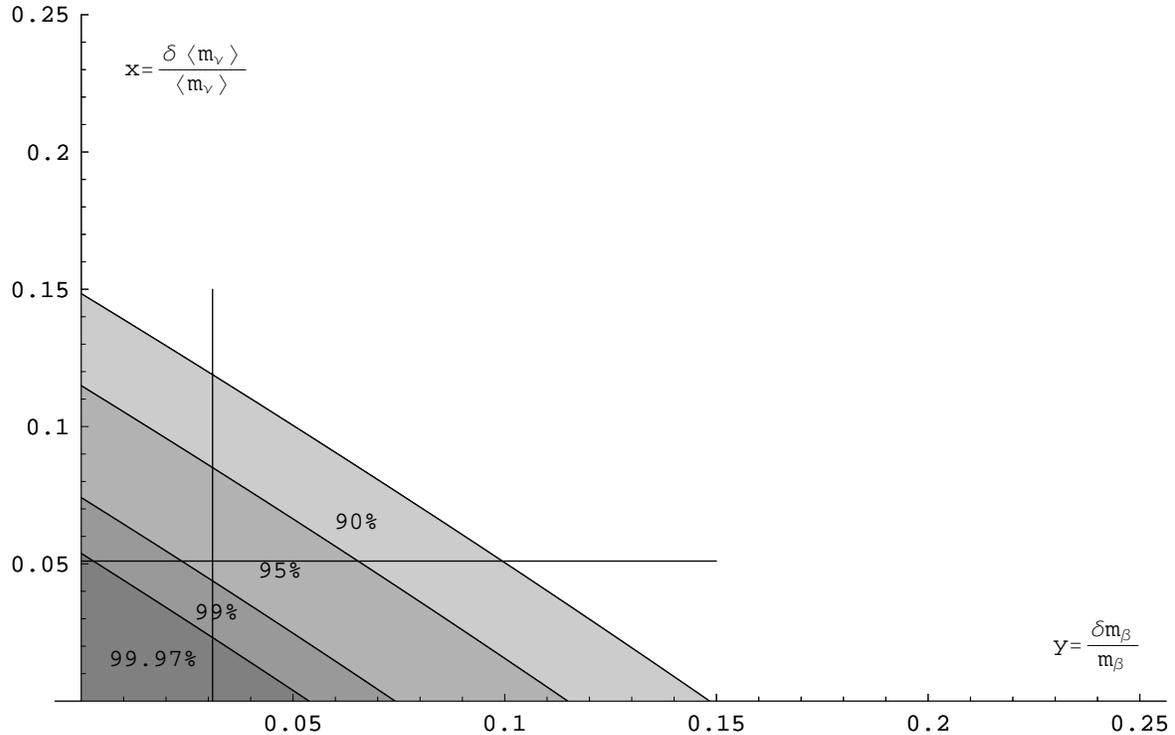}
\caption{The regions of relative error of $m_\nu$ versus
$m_\beta$, where CP is violated with confidence level (CL) equal
to: 90\%, 95\%, 99\% and 99.97\% (3$\sigma$). To find CP symmetry
breaking, $\langle m_\nu\rangle$ and $m_{\beta}$ should be determined
with extremly difficult to reach precision.} \label{fig5}
\end{figure}
\par More careful analysis, taking into account the present precision of
the mixing angle determination \cite{Gonzales} can give a region
in the $\left( \langle m_\nu\rangle, m_\beta\right)$ plane where CP
violation can be probe with various CL. The regions of relative
errors $\frac{\Delta m_\nu}{m_\nu}$ and $\frac{\Delta
m_\beta}{m_\beta}$ for which CP violation could be seen are
presented in Fig. 5. We see that even for 90\% CL the $x$
parameter should be smaller than $x<0.15$, so it is completely out
of reach with present experimental and theoretical possibilities.
\par How a better determination of the
$\theta_{12}$ and $\theta_{13}$ mixing angles affects the $x$ and
$\Delta m_\beta$ uncertainties? Let us assume that during next
years the precision of experiments will be strongly improved. Let
us also assume that the best values of mixing angles will not
change but only precision will be much better:
\begin{enumerate}
\item The KamLAND and Borexino experiments determine the solar
mixing angle with precision $\sin^2\theta_{12}\approx 0.28\pm
0.01$ \cite{Barger5} \item The IHF-Kamioka neutrino experiment or the
future neutrino factories \cite{Itow} will measure the $\theta_{13}$
with the precision $\Delta \theta_{13}=0.01$ (so
$\sin^2\theta_{13}= 0.005\pm 0.0001$).
\par \noindent End assume finally that:
\item Weak lensing of galaxies by large scale structure together
with CMB data measure the sum of neutrino masses
$\sum=m_1+m_2+m_3$ to an uncertainty of $0.04\: eV$. So we can
expect that each individual mass is known with the precision
$\Delta m_{\beta}= 0.015\:eV$ \cite{Hu}.
\end{enumerate}
Now from Eq.(\ref{xy}) we get the required precision of $\Delta
m_{\beta}$ and $\Delta \langle m_{\nu}\rangle$
\begin{equation}\label{xy2}
x,\:y<0.36.
\end{equation}
\begin{figure}
\includegraphics[width=\textwidth]{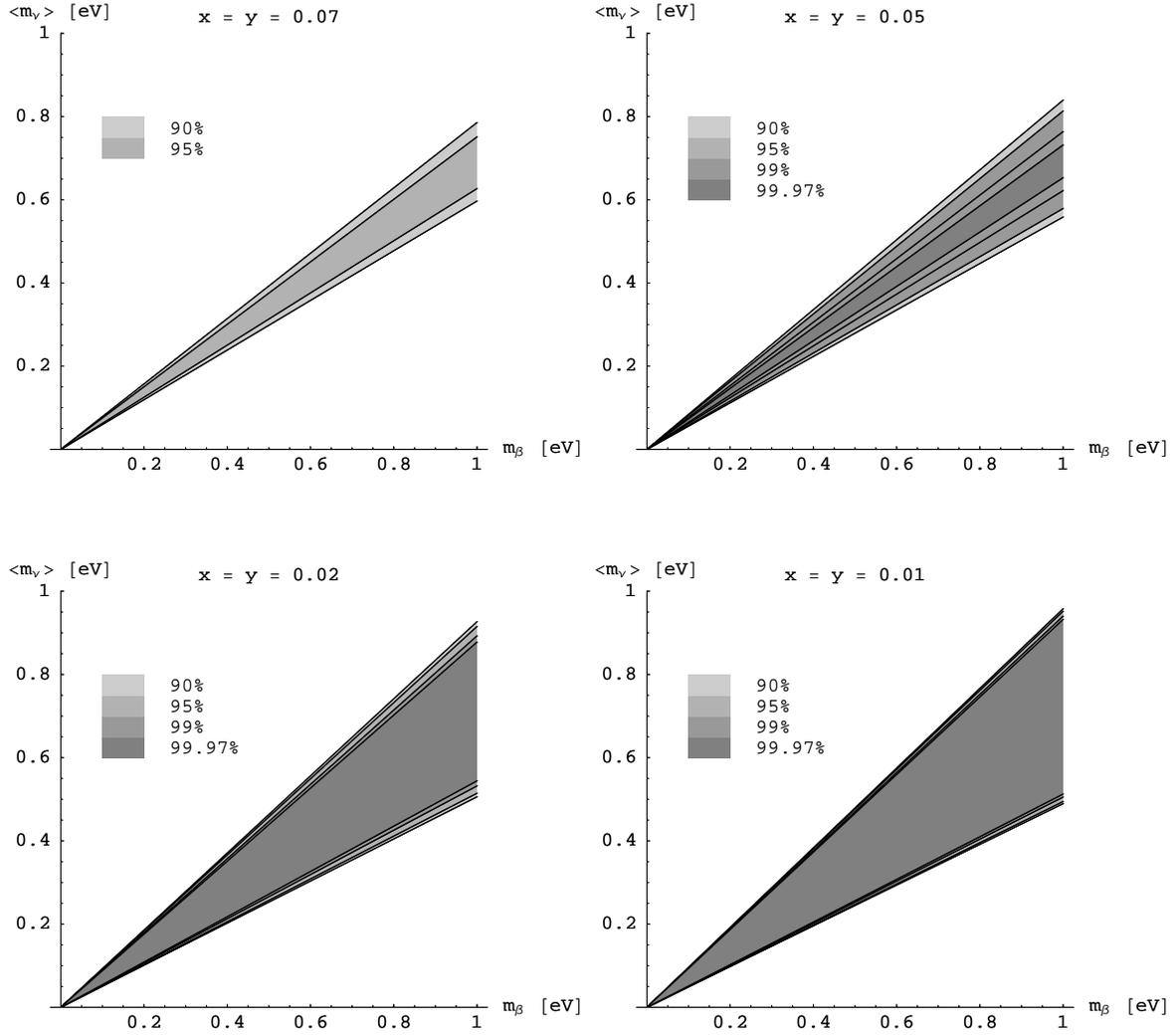}
\caption{Regions in the $\langle m_\nu\rangle\Leftrightarrow
m_\beta$ plane where CP symmetry is broken with various CL for
equal relative errors of $\langle m_\nu\rangle$ and $m_\beta$
(x=y).} \label{fig6}
\end{figure}
In Fig. 4  we present for this value of $x$ a necessary
precision of  NME for different relative errors of the
$T(^{76}Ge)$ measurements. If the last estimation of NME is
confirmed $(a\thickapprox 1.4)$ and the decay lifetime of
$^{76}Ge$ is found with $x_T\leq 0.5$ then necessary precision of
$\langle m_\nu\rangle$ will be obtained. Such a scenario is not only a
pure fantasy. More precise estimation will give a region of
$\langle m_\nu\rangle$ and $m_\beta$ where a probe of CP
violation could be possible (Fig. 6). We have assumed the same
relative uncertainties for $\langle m_\nu\rangle$ and $m_\beta$
$(x=y)$. For $x=y=0.07$ there is no region where CP could be find
with $CL>99\%$. This region appears if $x=y$ are smaller.
\begin{figure}
\includegraphics[width=\textwidth]{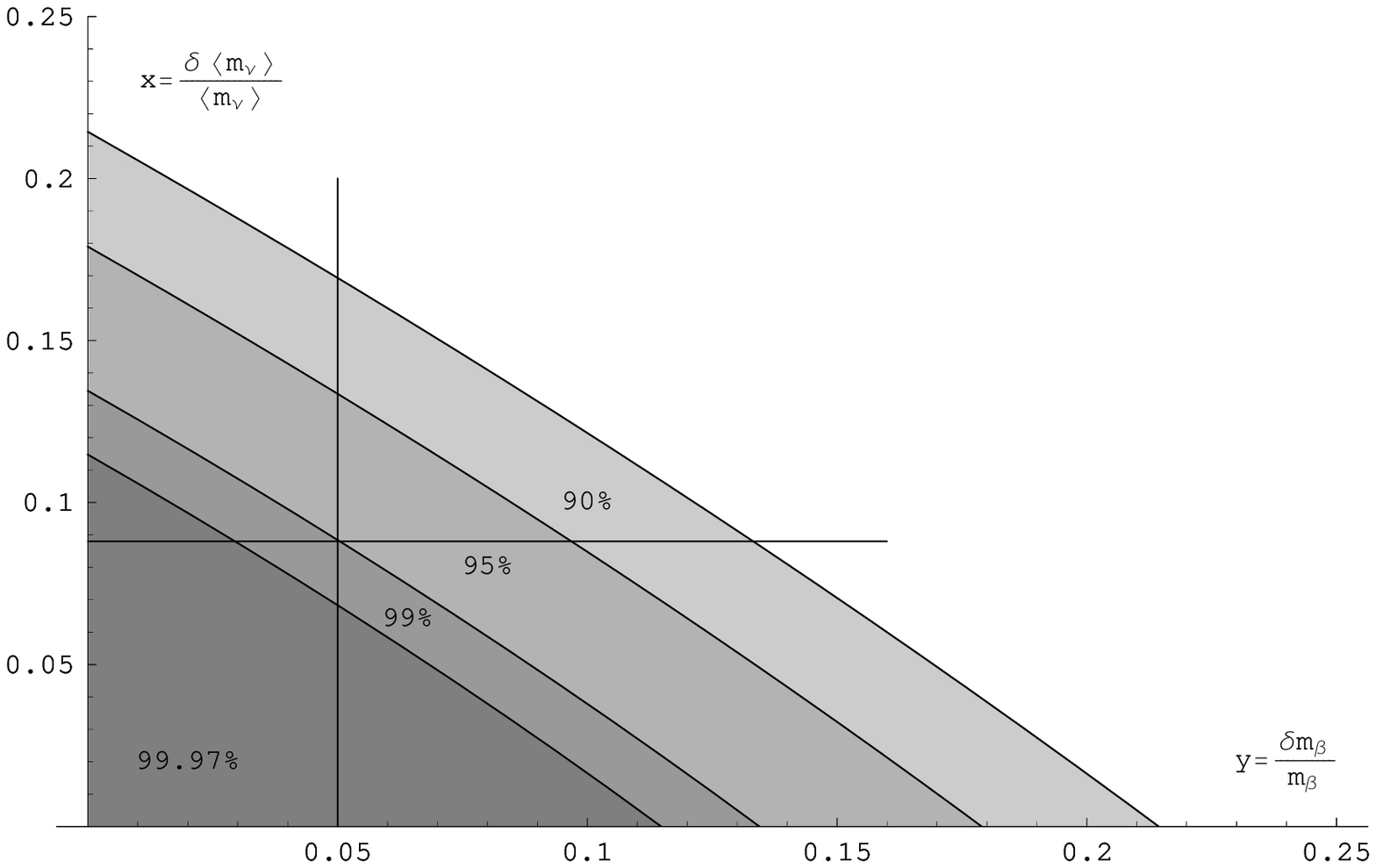}
\caption{The regions of relative errors of $\langle m_\nu\rangle$
versus $m_\beta$ where CP is violated with CL equal to: 90\%,
 95\%, 99\% and 99.97\% ($3\sigma$). If $y=0.05$ then to determine CP
 violation, at $99\%$ CL we have to know the effective Majorana
 mass with precision $x=0.09$.}
\label{fig7}
\end{figure}
 In Fig. 7 a region of $x$ and $y$ relative errors is presented for
a given level of CL. We can see that if we want to probe CP violation
with $CL\thickapprox90\%$ $x$ must
be smaller than $x\leq 0.22$ for very well determined $m_\beta$
($y\rightarrow0$) and vice versa, $y\leq 2$ for $x\rightarrow0$. Correlations
between quantities give
more stringent requirements for relative errors (see Eq.
(\ref{xy2})). We can see from Fig.4 that to get $x\sim 0.1$, parameter
$a$ must be smaller than $1.3$ and $x_T$ better than $10\%$.
Knownledge of NME on a $30\%$ level has been postulated recently \cite{Zuber}

\section{Conclusions}
\par From presented estimations it follows that measurement for CP violation
for Majorana neutrinos in neutrinoless double beta decay could be
possible for almost degenerate spectrum of their masses ($m_\beta>0.1$ eV).
However, several conditions should be satisfied:
\begin{enumerate}
\item oscillation mixing angles should be measured with better
precision e.g. $\Delta (\sin\theta_{13}\thickapprox0.01$ and
$\Delta(\sin\theta_{12} \thickapprox0.1)$ which are within the future
experimental range (see e.g.\cite{Barger5, Itow}).
 \item absolute neutrino masses $m_\beta$ should be measured with precision
 $\Delta m_\beta\thickapprox0.02$ eV with the central value in the range
 $m_\beta>0.15$ eV, which is also not a fully fantastic dream \cite{Hu}.
\item neutrinoless double beta decay is discovered and the decay
lifetime $T$ is measured with precision better than $10\%$. It is
difficult to say at the moment anything about the future precision of
T. If we give a credit to the last Heidelberg group news about
$(\beta\beta)_{0\nu}$ decay of $^{76}Ge$, then the error of $T$ is
much higher. They derived from the full data taken until May 2003
that \cite{Klapdor2}
\begin{equation}
T(^{76}Ge)=(0.69-4.18)\times10^{25}\;\mathrm{y},
\end{equation}
with the best value $T(^{76}Ge)=1.19\times10^{25}$ y, so the
relative error $x_T=\frac{\Delta T}{T}\sim 2.9$. To get $x_T<0.1$
will be probably a very difficult task. \item nuclear matrix
elements of decaying isotopes are calculated with much better
precision.
Future uncertainties for $a=\frac{b_{max}}{b_{min}}$ should be smaller
than $a<1.3$. During the last years some improvement in NME calculation has
been obtained. The last result where $a\approx1.4$ has been presented is a
very good step forward \cite{Rodin}. The model of NME calculation can also
be tested via comparison of the results of calculation for three (or more)
nuclei with experimental data \cite{Bilenky4,Bilenky5,Bilenky6}. This test
can be acomplished if $(\beta\beta)_{0\nu}$ decay of several nuclei is
observed. \item there should be
independent information about a full mechanism of the
$(\beta\beta)_{0\nu}$ decay. We should know that two electrons are
produced by two W-bosons and Majorana neutrino exchange virtual process.
Any other mechanism should give negligible contribution to
the neutrinoless electrons production. The future LHC data and
observation of other lepton violating processes give some chance to clarify this issue.
\cite{Cirigliano}.
\end{enumerate}
\begin{acknowledgments}
 This research was supported by Polish Ministry of Science under
Grant 1 P03 B 049 26.
\end{acknowledgments}


\bibliography{bib}

\end{document}